\documentclass[conference]{IEEEtran}
\IEEEoverridecommandlockouts

\usepackage{cite}
\usepackage{amsmath,amssymb,amsfonts}
\usepackage{algorithmic}
\usepackage{graphicx}
\usepackage{textcomp}
\usepackage{xcolor}
\usepackage{booktabs}
\usepackage{makecell}
\usepackage{balance}
\def\BibTeX{{\rm B\kern-.05em{\sc i\kern-.025em b}\kern-.08em
    T\kern-.1667em\lower.7ex\hbox{E}\kern-.125emX}}
\begin{document}

\title{WIP: Enhancing Game-Based Learning with AI-Driven Peer Agents\\
\thanks{This work has been accepted for presentation at IEEE Frontiers in Education Conference 2025.%
    The final published version will be available via IEEE Xplore.}
}

\author{
Chengzhang Zhu\textsuperscript{1}, Cecile H. Sam\textsuperscript{2}, Yanlai Wu\textsuperscript{1}, Ying Tang\textsuperscript{1\dag} \\
\textsuperscript{1}Department of Electrical and Computer Engineering, Rowan University, USA \\
\textsuperscript{2}Department of Educational Leadership, Rowan University, USA \\
Email: zhuche95@students.rowan.edu, sam@rowan.edu, wuyanl37@rowan.edu, tang@rowan.edu
}

\maketitle

\begin{abstract}This work-in-progress paper presents SPARC (Systematic Problem Solving and Algorithmic Reasoning for Children), a gamified learning platform designed to enhance engagement and knowledge retention in K-12 STEM education. Traditional approaches often struggle to motivate students or facilitate deep understanding, especially for complex scientific concepts. SPARC addresses these challenges by integrating interactive, narrative-driven gameplay with an artificial intelligence peer agent built on large language models. Rather than simply providing answers, the agent engages students in dialogue and inquiry, prompting them to explain concepts and solve problems collaboratively. The platform’s design is grounded in educational theory and closely aligned with state learning standards. Initial classroom pilots utilized a multi-method assessment framework combining pre- and post-tests, in-game analytics, and qualitative feedback from students and teachers. Preliminary findings indicate that SPARC significantly increases student engagement, with most participants reporting greater interest in STEM subjects and moderate gains in conceptual understanding observed in post-test results. Ongoing development focuses on refining the AI agent, expanding curriculum integration, and improving accessibility. These early results demonstrate the potential of combining AI-driven peer support with game-based learning to create inclusive, effective, and engaging educational experiences for K-12 learners.
\end{abstract}


\section{Introduction}
This research-to-practice paper describes the implementation and initial evaluation of SPARC (Systematic Problem Solving and Algorithmic Reasoning for Children), an innovative, AI-powered gamified learning platform designed to enhance K-12 STEM education through immersive, interactive experiences.

Traditional K-12 STEM education often faces significant challenges. A major challenge is maintaining student engagement, as traditional teaching methods typically place students in passive learning roles, limiting their motivation and interest. In addition, traditional classroom methods often struggle to ensure effective knowledge transfer\cite{Tranbad} due to the limitations of the scenarios, which results in students not being able to apply their theoretical knowledge to actual real-world scenarios.

To bridge these educational gaps, the SPARC project introduces a narrative-driven, interactive learning environment. Central to SPARC’s current research are AI-powered peer agents. They function not just as passive mentors\cite{Ryan} on narratives but as active peers. In our design, they can prompt and guide students through structured dialogue and inquiry-based sessions\cite{TraTeach}. This unique integration of AI peer agents aims to increase student engagement, critical thinking, and conceptual understanding by transforming passive learning into a dynamic, student-led inquiry.


In the following sections, we outline the theoretical underpinnings that guide this practice, detail our implementation process, discuss preliminary findings, and highlight the impact and future directions of this innovative educational approach.

\section{Background and Motivation}

Educational research is increasingly supporting the use of interactive and personalized approaches to address traditional limitations. Gamified learning environments that leverage intrinsic motivators like curiosity, challenge, and enjoyment have demonstrated significant potential to enhance student motivation and engagement compared to traditional methods\cite{Gamifi}\cite{funlearn}. Specifically, games promote positive learning experiences, provide immediate feedback, encourage problem-solving, and promote a sense of accomplishment - all factors that are linked to improved learning outcomes\cite{Gamelearn1}\cite{Gamelearn2}.

In addition, recent theoretical developments have emphasized the importance of peer learning and social interaction in educational settings. According to connectivism theory, students learn best when they actively make connections between concepts through collaborative communication and interactive resources with their peers\cite{connectivism}. Such interactions encourage students to articulate their understanding and promote deeper understanding and retention of knowledge\cite{Gameeva}.

Artificial intelligence (AI), like large language models, adds another important dimension to personalized education. AI-powered systems can adaptively tailor content and feedback to the individual needs of each learner, effectively delivering personalized learning experiences at scale\cite{adaptive}\cite{RyanModel}. If AI-powered peer agents are integrated into a gaming environment, they can provide a unique opportunity for active dialogue and personalized support\cite{Wang2024}, combining the motivational benefits of gaming with the adaptive benefits of AI.

Driven by these insights, the SPARC project aims to leverage AI-powered peer agents in gamified STEM learning experiences that specifically target the engagement and conceptual understanding of underserved student populations. This work-in-progress explores the early implementation and preliminary effects of this innovative approach to address gaps in traditional K-12 STEM education.



\section{Research-to-Practice Framework}


The design of SPARC is grounded in well-established learning theories that emphasize active, contextualized, and socially mediated learning. From a constructivist perspective, students learn most effectively by constructing meaning through hands-on experience and reflection, rather than passively receiving information\cite{Constructivism}. Informed by this principle, SPARC integrates learn-by-doing activities and problem-based challenges to promote conceptual understanding within meaningful contexts.

In addition, social learning theories\cite{zhou2015educational}, particularly those emphasizing peer collaboration and situated interaction, influence the game's design through features such as AI-powered peer agents and cooperative gameplay tasks. These elements promote articulation, modeling, and mutual support—critical mechanisms for deeper learning in collaborative environments.

Cognitive learning theories also play a vital role by informing the system's use of feedback, memory reinforcement, and metacognitive prompts\cite{Cog}. For instance, SPARC's peer agents actively prompt students to explain their reasoning, a practice known to enhance both retention and conceptual clarity.

Together, these complementary theoretical foundations guide the core mechanics of SPARC, ensuring that gameplay is not only engaging but also aligned with empirically supported instructional strategies.

\subsection{Design Principles in SPARC}
Building on these theories, SPARC applies core principles of educational design to create activities that are both engaging and instructional. The platform’s content is fully aligned with the New Jersey Student Learning Standards for Science (NJSLS-S).

\subsubsection{Immersive Narrative Context}
\begin{figure}[htbp]
\centering{\includegraphics[width=\columnwidth]{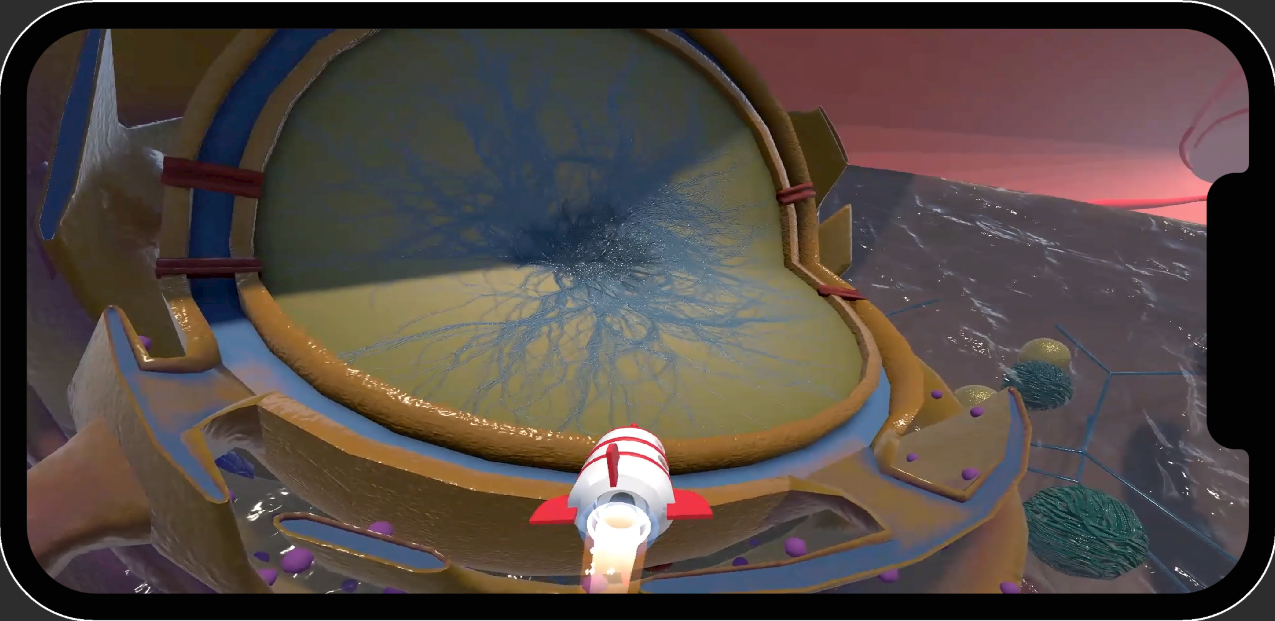}}
\caption{Rocket exploration.}
\label{figRoc}
\end{figure}

SPARC takes a story-driven approach to building a learning framework. Inspired by the Magic School Bus, the game's storyline shrinks the player to explore the human body in a rocket ship, interacting with cells and organs along the way, as is shown in Fig.\ref{figRoc}. By embedding the curriculum within the adventure, SPARC utilizes storytelling to capture attention and provide meaning to the tasks that guide the journey.

Although SPARC’s storyline includes imaginative elements like exploring the human body in a rocket ship, all educational content, dialogue, and environments are thoroughly reviewed by biology experts. This ensures that engaging, real-world connections are made without sacrificing scientific accuracy or introducing misconceptions.

\subsubsection{AI Peer Agent Interaction}
\begin{figure}[htbp]
\centering{\includegraphics[width=\columnwidth]{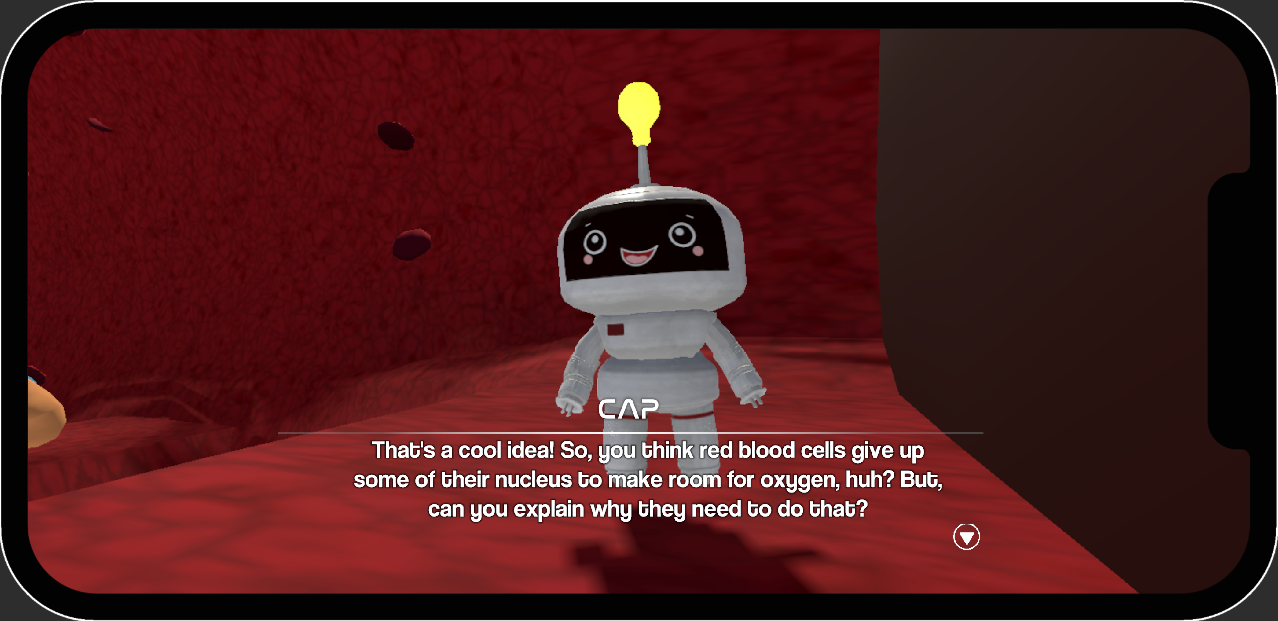}}
\caption{Agent interaction.}
\label{figagent}
\end{figure}

The game features an intelligent companion agent that serves as a conversational partner, rather than a traditional Non-Player Character (NPC). This AI-driven character accompanies the student through the game, guiding the storyline and the player's exploration. As the game progresses, the agent builds trust with the player and reinforces the player's memory by prompting and asking questions at appropriate times, as is shown in Fig.\ref{figagent}.

\subsubsection{Collaborative and Social Learning}
SPARC supports teamwork and individual exploration to mirror real-life learning environments. The game allows students to form teams (up to four) to work together to solve problems, share clues, and view collective and individual progress. Built-in chat and group goals encourage peer discussion and collaboration during gameplay, building communication skills and peer learning.
Meanwhile, independent puzzle-solving tasks allow for independent learning, ensuring that collaboration does not interfere with individual problem-solving practice. This balance helps accommodate different learning preferences and promotes social interaction as a learning vehicle.



\subsubsection{Gameplay Diversity and Engagement Systems}

To further encourage ongoing student participation, SPARC integrates a variety of engaging and replayable gameplay modules. For instance, the cardiovascular module includes a rhythm-based music game, and the oxygen transport activity is framed as a high-speed racing challenge. These interactive elements are designed to make learning enjoyable and to motivate students to return for repeated play. Additionally, SPARC features an achievement system that rewards students with cosmetic unlocks for reaching milestones in different play styles, as well as a global leaderboard that fosters friendly competition among players. The platform is continuously updated with new content and user-driven enhancements, ensuring that students have ongoing reasons to re-engage and explore new challenges. These mechanics transform SPARC from a one-time experience into a dynamic and lasting learning environment.

\subsection{Iterative Design and Curriculum Alignment}
An important aspect of translating research into practice is an iterative, user-centered development process. The SPARC team collaborated with K-12 teachers and students across multiple design cycles. Ongoing teacher feedback ensured that the game’s content remained closely aligned with NJSLS, leading to targeted curricular adjustments, such as adding Force and Motion to the Blood Circulation Journey. Each SPARC module is purposefully designed to support standards-based instruction, with current content focused on the circulatory system and new modules (e.g., digestive system) actively being developed. At the same time, student feedback during testing led to practical improvements in game mechanics and interface usability, such as customizable viewpoints and a reset feature, which addressed motion sickness in early prototypes.

\subsection{Peer Agent in Practice – Learning by Teaching}

Artificially intelligent peer agents (“CAP”) are among SPARC’s most innovative features, translating educational research on peer tutoring and teachable agents into engaging game mechanics. Rather than lecturing, CAPs interact with students as peers—asking questions, seeking clarification, and prompting learners to explain concepts, which deepens understanding by placing students in a teaching role. This approach, rooted in evidence that teaching others enhances learning, makes classroom interactions both motivational and instructional. Teachers and students report that CAPs create a more immersive and educational game environment.

Technically, the CAP agent is implemented as a modular system based on a locally deployed Llama3 language model, which has been fine-tuned for educational use.
All student behaviors, including navigation, task completion, and dialogue, are logged with metadata to support both learning analytics and the agent’s internal memory module. 
This memory tracks both short-term and long-term context, indexing events by time, importance, and relevance to each student’s evolving knowledge map, and calculates a curiosity index to enable adaptive prompts. 
The modular design also allows for future integration of reinforcement learning from human feedback and emotion recognition, further enhancing the agent’s ability to deliver personalized support. 


\section{Method of Assessment}
To evaluate SPARC’s educational impact and practical implementation, we adapted a multi-method assessment framework\cite{patton2015qualitative} that combines quantitative and qualitative measures of learning outcomes, interaction with AI-driven peer agents, and student engagement.

\subsection{Pre- and Post-test Assessments}
Student learning gains were measured through pre- and post- tests administered before and during gameplay. These assessments were teacher-informed and vetted. The assessments covered key STEM topics and allowed for direct comparison of conceptual understanding. Results were analyzed to determine knowledge improvement attributable to SPARC.

\subsection{In-Game Interaction Analytics}
In-game analytics captured metrics such as module completion rates, frequency of interactions, time-on-task, and engagement patterns. These behavioral data informed the ongoing optimization of game mechanics and adaptive feedback. Currently, data were analyzed post hoc, but future iterations will incorporate real-time analytics for adaptive in-game support and deeper agent integration.

\subsection{Student-Agent Dialogue Analysis}
Qualitative analysis of student-agent dialogues focused on conceptual accuracy, depth of explanation, and effective use of vocabulary. Transcripts were systematically coded to assess how student interactions with the peer agent reinforce STEM learning and address misconceptions.

\subsection{Student Feedback Surveys}
Surveys collected descriptive feedback from students covering perceptions of learning, motivation, usability, and classroom integration. A thematic analysis of qualitative responses helped identify recurring strengths, areas for improvement, and practical considerations for implementation.

\subsection{Student Focus Groups}
To complement the student surveys, 4 focus groups (ranging between 3-6 students) were conducted after gameplay to gain a fuller understanding of their experience. Students spoke about preferred modes of learning in relation to the game content, suggestions for improvements, and how the game fits amongst their overall videogame experiences.

By integrating these assessment methods and using fieldnotes and observations for triangulation \cite{phillippi2018fieldnotes},  we obtained a comprehensive understanding of SPARC’s effectiveness and continuously inform further development.

\section{Preliminary Findings}


Although the SPARC project is still ongoing, multiple rounds of classroom pilots have provided valuable insights into its educational potential. Specifically, pilot implementations conducted in 2024 and 2025 allowed for staged data collection across both subjective feedback and objective performance measures.

\subsection{Student Perceptions from 2024 and 2025 Pilots}

Two rounds of SPARC deployment gathered perception data through student surveys. The first pilot in 2024 involved 56 students, while a follow-up pilot in 2025 collected responses from 54 students. These surveys focused on student-reported learning, enjoyment, perceived usefulness, and comparison with traditional instruction.





Across both pilots, the results indicate strong student engagement and positive perceptions of the platform. In the 2024 early version pilot, over 80\% of students reported increased interest in STEM fields, and 89\% found SPARC to be effective or superior to traditional instruction. Approximately 72\% believed that they would recommend this game to their friends.

In the recent survey for the 2025 version shown in Table~\ref{tab:sparc_summary}, a total of 54 students (29 identified as boys and 25 identified as girls ) reported a moderate to high level of learning from the game (M = 3.81 on a 5-point scale). Regarding problem-solving skills, somewhat was the most frequently reported response (51.9\%), followed by a lot (20.4\%). Most students described their overall gameplay experience positively, with 42.3\% reporting it as enjoyable, 30.8\% as neutral, and none reporting it as not enjoyable. Additionally, the majority indicated that they believed the game could help other students their age learn about the topics, and many expressed interest in learning more about how the human body works after playing.

\begin{table}[htbp]
\centering
\caption{Summary of Student Perceptions of SPARC Game Experience}
\label{tab:sparc_summary}
\setlength{\tabcolsep}{4pt}
\begin{tabular}{lccccc}
\toprule
\makecell[l]{Survey Item} & 1 & 2 & 3 & 4 & 5 \\
\midrule
\makecell[l]{Overall learning \\ from SPARC} & 0.0\% & 1.9\% & 33.3\% & 46.3\% & 18.5\% \\
\makecell[l]{Improved problem- \\ solving skills} & 5.6\% & 14.8\% & 51.9\% & 20.4\% & 7.4\% \\
\makecell[l]{Overall enjoyment \\ of the gameplay experience} & 0.0\% & 3.8\% & 30.8\% & 42.3\% & 23.1\% \\
\makecell[l]{SPARC compared to \\ traditional assignments} & 1.9\% & 3.7 & 33.3\% & 44.4\% & 16.7\% \\
\bottomrule
\end{tabular}

\vspace{0.5em}
\begin{minipage}{0.9\linewidth}
\footnotesize
\textit{Note: Response options on a 5-point Likert scale. For Items 1–3, 1 = Not at all, 5 = A significant amount. For Item 4, 1 = Much worse than traditional assignments, 5 = Much better. N = 54.}
\end{minipage}
\end{table}



\subsection{Learning Outcomes and Cognitive Gains}


In addition to survey data, the 2025 pilot included pre- and post-test assessments to evaluate conceptual understanding. A total of 60 students completed the pre-test before gameplay, and 44 of them provided valid post-test data following their interaction with the SPARC modules. 

The modules differed primarily in interaction style and instructional methods, specifically regarding whether students interacted with an AI peer agent employing adaptive LLM-generated inquiry prompts (Narrative story modules), passive instructional videos (Passive Video Instruction Modules), or direct, hands-on gameplay tasks (Interactive Gameplay Modules). Table~\ref{tab:sparc_prepost} summarizes accuracy scores across these types.

\begin{table}[ht]
\centering
\caption{Pre- and Post-test Accuracy (\%) by Module Type (2025)}
\label{tab:sparc_prepost}
\begin{tabular}{lccc}
\toprule
Module Type & Pre-test (\%) & Post-test (\%)\\
\midrule
Agent-based Module(LLM Inquiry) & 39.2 & 45.7\\
Agent-based Module(No LLM Inquiry) & 39.2 & 24.4\\
Passive Video Instruction Modules & 50.0 & 28.3\\
Interactive Gameplay Modules & 47.3 & 59.2\\
Overall & 46.6 & 49.7\\
\bottomrule
\end{tabular}
\vspace{0.8em}

\begin{minipage}{0.9\linewidth}
\footnotesize
\textit{Note: N = 60 for pre-test; N = 44 for post-test.}
\end{minipage}
\end{table}

Notably, students who completed agent-supported (LLM Inquiry) modules showed meaningful improvements in post-test scores compared to other groups, despite the greater conceptual difficulty of these modules. Compared to other modules, the LLM Inquiry modules also featured deeper and more cognitively demanding questions, which may have contributed to the observed gains. In contrast, modules that used only video showed lower post-test scores, suggesting that passive formats were less effective for learning new concepts. We also noticed that some pre-test scores were higher than expected because students discussed answers or looked them up online. This may help explain the drop in post-test scores, especially for the video-based modules.

Taken together, these findings suggest that interactive, agent-driven learning environments better support student understanding of complex topics than either non-agent or video-based modules. Future pilots will further strengthen experimental control and incorporate direct comparison groups to more rigorously evaluate learning gains attributable to SPARC.

\subsection{Positive Feedback from Students}
Analyses of student qualitative responses in both open-ended questions and focus groups revealed overwhelmingly positive attitudes toward  SPARC’s interactive learning experience. Aligning with survey results, students reported that the SPARC experience was a novel learning experience that they found to be a good complement to teacher instruction. In focus groups, students wanted more interactive activities like SPARC, but also valued the role that teachers played in providing the foundational information. Students expressed high engagement stating, "The game kept me focused and helped me learn more effectively," and "I found it exciting to explore and understand different parts of the human body."


\section{Challenges and Lessons Learned}
Through the initial implementation of SPARC, several practical and technical challenges have been identified, providing critical lessons for future improvements.

\subsection{Educational Implementation Challenges}
Classroom implementation revealed notable challenges:
\begin{itemize}
    \item Limited device availability and inconsistent technical infrastructure (e.g., unreliable Wi-Fi and shared devices) impeded stable implementation.
    \item Technical difficulties resulted in incomplete data collection, impacting the accuracy and completeness of assessments.
\end{itemize}

To address these issues, future releases of SPARC will include an offline mode that enables students to access core learning modules without a continuous internet connection. The platform is also being optimized for wider compatibility with devices, including tablets, smartphones, and shared classroom devices, to maximize accessibility in a variety of educational settings. In addition, a technical checklist and troubleshooting guide will be provided to teachers, allowing for rapid diagnosis and resolution of common connectivity or hardware issues. These improvements are intended to minimize technical disruptions and ensure equitable, consistent access for all students.

\subsection{Technical and User Experience Challenges}
First, Initial user interface (UI) designs presented usability difficulties for some students, resulting in confusion and navigation challenges. Second, although generally effective, AI-driven peer agents required further improvement in real-time responsiveness, accuracy, and naturalness of interactions.

In such a case, future research and development will focus on refining user interface interactions and enhancing AI models, aiming to provide more intuitive and seamless student experiences.

\section{Future Work and Implications}
Moving forward, we will focus on expanding assessments and refining the SPARC system to address these identified challenges. Planned next steps include larger-scale trials in more classrooms and schools to collect robust data, validate initial positive trends, and assess knowledge growth and retention with greater confidence. With increased sample size and study duration, we can better evaluate learning gains and fine-tune the platform accordingly.

The AI-driven features of SPARC, especially the peer agent, will be further enhanced to offer more personalized and adaptive support to diverse learners. We are developing the agent module as a stand-alone, modular component that can be flexibly integrated into a wide range of educational games and platforms, making it adaptable to any subject area or digital environment. This modularity aims to maximize the broader applicability and impact of AI-supported peer learning in education.

Another focus is on deepening curriculum integration. We continue to work with educators to align SPARC content with school standards and provide guidelines and training for seamless classroom integration\cite{Cui2023}. By improving accessibility, AI adaptivity, and curriculum fit, SPARC aims to extend its impact well beyond initial pilots.

Ultimately, gamified AI-powered platforms like SPARC, especially through accessible mobile technology, have the potential to democratize STEM education for K-12 learners. We anticipate that the lessons learned from SPARC will help guide the design of future AI-powered educational tools and inspire the next generation of scientists and engineers.

\section{Conclusion}
This work-in-progress paper presents the initial implementation and evaluation of SPARC, a gamified learning platform enhanced by AI-driven peer agents. Early results suggest that SPARC can substantially boost student engagement and interest in STEM through interactive dialogue and immersive gameplay. While practical and technical challenges remain, the platform demonstrates strong potential for advancing K-12 education with AI-enhanced gamification. Continued development and larger-scale studies will further refine SPARC and support its broader impact on inclusive STEM learning.

\section*{Acknowledgment}

 This work is supported in part by Google Academic Research Award and Rowan University-Rutgers-Camden Board of Governors.

\balance
\bibliographystyle{IEEEtran}

\bibliography{ref}

\end{document}